\documentclass[prc,twocolumn,superscriptaddress,superscriptfootnote]{revtex4}
\usepackage{amssymb}

\usepackage{amsmath,amsfonts,amssymb,epsfig,color}

\begin{document}

\title{The proton-neutron symplectic model of nuclear collective motions}
\author{H. G. Ganev}
\affiliation{Joint Institute for Nuclear Research, Dubna, Russia}

\setcounter{MaxMatrixCols}{10}

\begin{abstract}
A proton-neutron symplectic model of collective motions, based on
the non-compact symplectic group $Sp(12,R)$, is introduced by
considering the symplectic geometry of the two-component
many-particle nuclear system. The possible classical collective
motions are determined by different dynamical groups that can be
constructed from the symplectic generators. The relation of the
$Sp(12,R)$ irreps with the shell-model classification of the basis
states is considered by extending of the state space to the direct
product space of $SU_{p}(3) \otimes SU_{n}(3)$ irreps, generalizing
in this way the Elliott's $SU(3)$ model for the case of
two-component system. The $Sp(12,R)$ model appears then as a natural
multi-major-shell extension of the generalized proton-neutron
$SU(3)$ scheme which takes into account the core collective
excitations of monopole and quadrupole, as well as dipole type
associated with the giant resonance vibrational degrees of freedom.
Each $Sp(12,R)$ irreducible representation is determined by a
symplectic bandhead or an intrinsic $U(6)$ space which can be fixed
by the underlying proton-neutron shell-model structure, so the
theory becomes completely compatible with the Pauli principle. It is
shown that this intrinsic $U(6)$ structure is of vital importance
for the appearance of the low-lying collective bands with both the
positive and negative parity. The full range of low-lying collective
states can then be described by the microscopically based intrinsic
$U(6)$ structure, renormalized by coupling to the giant resonance
vibrations.

\end{abstract}
\maketitle
PACS {21.60.Fw, 21.60.Ev}

\section{Introduction}

Symmetry is an important concept in physics. In finite many-body
systems, it appears as time reversal, parity, and rotational
invariance, but also in the form of dynamical symmetries
\cite{DG}-\cite{RW}. The standard symmetry approach allows the
construction of a Hamiltonian of a system under consideration which
is, or nearly so, invariant under a group of symmetry
transformations. Group theory then allows one to construct basis
states realizing the symmetry and explicit matrix elements for
physically interesting transition operators themselves classified by
the symmetry. Many properties of atomic nuclei have been
investigated using algebraic models, in which one obtains bands of
collective states which span irreducible representations of the
corresponding dynamical groups \cite{aans},\cite{Frank},\cite{RW}.

There is, however, another non-standard approach for exploiting the
symmetry by identifying first the generators of possible collective
motions and then the algebra they close under commutation
\cite{NA1}-\cite{NA4}. This reveals directly the physical content of
a certain algebraic model. The Hamiltonian of nuclear system is then
assumed to be a function of these operators. Along this line, some
algebraic models of collective motions in nuclei have been proposed
based on the algebras of $SL(3,R)$ \cite{sl3R}, $Rot(3)$
\cite{rot3}, $CM(3)$ \cite{cm3} and $Sp(6,R)$ \cite{RR1} groups,
respectively.

It is known that in formulating the nuclear many-body problem some
kinematical requirements should be satisfied by the nuclear wave
function \cite{Van74},\cite{Van76}. First, the wave function of the
nucleus should be realized microscopically, i.e. it should depend
upon all single particle variables -spacial and spin variables.
Secondly, the nuclear wave function should be
translationally-invariant. This means that the wave function of the
atomic nucleus, free from external fields, can be expressed as a
product of the plane wave, describing the center-of-mass motion, and
translationally-invariant wave functions, describing internal
properties of the free nucleus. The two conditions can be unified
into one and formulated as a requirement of the wave function to be
microscopically translationally-invariant. Third, the nuclear wave
function should preserve the observed integrals of motion (total
angular momentum, its third projection, etc.). An arbitrary wave
function fulfilling the above requirements is referred to as a
kinematically-correct wave function \cite{Van74},\cite{Van76}.

In the present paper we exploit further the non-standard symmetry
approach to nuclear collective motion and take into account
explicitly the proton-neutron degrees of freedom. Hence, we do not
use the isospin formalism and formulate the algebraic approach for
the two-component nuclear system. This allows to reveal some new
features and forms of collective excitations which are missing in
the microscopic theory of the one-component nuclear systems. We
propose a symplectic model, consistent with the proton-neutron
composite (granular) structure of the nucleus and appropriate mainly
for the description of different collective excitations in deformed
heavy mass even-even nuclei. The consideration of the symplectic
geometry shows that the $Sp(12,R)$ group appears as a dynamical
group of the collective excitations in the two-component many-body
nuclear system.

From the hydrodynamic perspective, the possible classical collective
flows are determined by different dynamical groups that can be
constructed from the symplectic generators of $Sp(12,R)$. To
quantize the model one has to construct the irreducible
representations of the $Sp(12,R)$ group, appropriate to the
many-particle system. The relation of the $Sp(12,R)$ irreps with the
shell-model classification of the basis states is considered by
extending of the state space to the direct product space of
$SU_{p}(3) \otimes SU_{n}(3)$ irreps, generalizing in this way the
Elliott's $SU(3)$ model \cite{Elliott58} for the case of
two-component system. The $Sp(12,R)$ model appears then as a natural
multi-major-shell extension of the generalized proton-neutron
$SU(3)$ scheme which takes into account the core collective
excitations of monopole and quadrupole, as well as dipole type
associated with the giant resonance vibrational degrees of freedom.
Each $Sp(12,R)$ irreducible representation is determined by a
symplectic bandhead or an intrinsic $U(6)$ space which can be fixed
by the underlying proton-neutron shell-model structure, so the
theory becomes completely compatible with the Pauli principle. It is
shown that this intrinsic $U(6)$ structure is of vital importance
for the appearance of the low-lying collective bands with both the
positive and negative parity. Then the full range of low-lying
collective states can be described by the microscopically based
intrinsic $U(6)$ structure, renormalized by coupling to the giant
resonance vibrations.

\section{The symplectic geometry}

In the microscopic nuclear theory the wave function should depend
upon all single particle variables $r_{1}, r_{2},\ldots ,r_{A}$. But
in order to avoid the problem of center-of-mass motion it is
convenient to use translationally-invariant variables from the very
beginning. The set of the translationally-invariant variables is
well known and is provided by the set of Jacobi vectors.
Additionally, nuclei consist of protons and neutrons. Thus, we
consider a two-component nuclear system consisting (after removing
of the center-of-mass) of $m=A-1$ particles and label the set of
Jacobi vectors, corresponding to protons and neutrons, by an
additional quantum number $\alpha = p,n$ (or $\alpha = 1,2$). The
latter extends the single particle configuration space to
$\mathbb{R}^{6}$. Then the Jacobi coordinates $x_{is}(\alpha)$ and
corresponding momenta $p_{is}(\alpha)$ of this $m$ particle system,
defined by the only nonzero commutator
\begin{equation}
\quad [x_{is}(\alpha),p_{jt}(\beta)]=i\hbar \delta_{ij}\delta_{st}
\delta_{\alpha\beta}, \label{commut}
\end{equation}
where $s,t=1,2....,m$, $i,j=1,2,3$, and $\alpha,\beta = p,n$, are
the elements of a $6m$-dimensional Heisenberg-Weyl Lie algebra
$hw(6m)$. The Hermitian quadratic expressions in the coordinates and
momenta
\begin{eqnarray}
&&x_{is}(\alpha)x_{jt}(\beta), \notag\\
&&x_{is}(\alpha)p_{jt}(\beta)+p_{jt}(\beta)x_{is}(\alpha), \notag\\
&&p_{is}(\alpha)p_{jt}(\beta), \label{Generators}
\end{eqnarray}
close under commutation the symplectic group $Sp(12m,R)$, which is
the full dynamical group of the system with $6m$ degrees of freedom.

The problem of $6m$ degrees of freedom defined in the many-particle
Hilbert space can be associated with a definite irrep of the
dynamical symmetry group $Sp(12m,R)$. The latter is also a dynamical
symmetry group of the $6m$-dimensional harmonic oscillator that
provides a complete set of states for the many-body problem.
Clearly, a large class of many-body Hamiltonians can be written in
terms of elements in the enveloping algebra of $Sp(12m,R)$. However,
it was proved that the collective effects are associated with
operators that are scalar in $O(m)$ \cite{Van76}$-$\cite{GC}. Then,
the collective part of the Hamiltonian is obtained by projecting the
latter on a definite $O(m)$ irrep \cite{Van76}$-$\cite{GC}
associated with the $m$ Jacobi vectors in the configuration space
$\mathbb{R}^{6m}$, where $m=A-1$ and $A$ is equal to the total
number of nucleons in the system.

The group $Sp(12m,R)$ among its subgroups has
\begin{equation}
Sp(12m,R) \supset Sp(12,R) \otimes O(m), \label{SpO}
\end{equation}
whose generators are obtained from (\ref{Generators}) in standard
way by means of a contraction. The infinitesimal operators of the
$O(m)$ group have the well-known antisymmetrized form
\begin{equation}
\quad
L_{st}=\sum_{i,\alpha}\biggl(x_{is}(\alpha)p_{it}(\alpha)-x_{it}(\alpha)p_{is}(\alpha)\biggr).
\label{OnGen}
\end{equation}
For $Sp(12,R)$ there are 78 Hermitian generators which are given by
the following one-body operators:
\begin{eqnarray}
&&Q_{ij}(\alpha,\beta)=\sum_{s=1}^{m}x_{is}(\alpha)x_{js}(\beta),
\label{Sp12a} \\
&&S_{ij}(\alpha,\beta)=
\sum_{s=1}^{m}\biggl(x_{is}(\alpha)p_{js}(\beta)+p_{is}(\alpha)x_{js}(\beta)\biggr),
\label{Sp12b} \\
&&L_{ij}(\alpha,\beta)=\sum_{s=1}^{m}\biggl(x_{is}(\alpha)p_{js}(\beta)-x_{js}(\beta)p_{is}(\alpha)\biggr),
\label{Sp12c} \\
&&T_{ij}(\alpha,\beta)=\sum_{s=1}^{m}p_{is}(\alpha)p_{js}(\beta).
\label{Sp12d}
\end{eqnarray}
As can be seen from (\ref{Sp12a})$-$(\ref{Sp12d}), the $Sp(12,R)$ is
generated by those bilinear operators which are invariant under
$O(m)$, and hence its generators commute with those of $O(m)$. Thus,
the two groups $Sp(12,R)$ and $O(m)$ are complementary
\cite{Filippov81},\cite{MQ},\cite{Van88},\cite{Rowe2012} within the
$Sp(12m,R)$ irrep, i.e. there is a relationship between their
irreps. This means that the quantum numbers labeling the $Sp(12,R)$
irrep $\langle \omega \rangle \equiv \langle
\omega_{1}+\frac{1}{2}m,\ldots,\omega_{6}+\frac{1}{2}m \rangle$
within the $Sp(12m,R)$ irreps $\langle \frac{1}{2}^{6m} \rangle$ or
$\langle \frac{1}{2}^{6m-1}\frac{3}{2} \rangle$ label also the
$O(m)$ irrep $(\omega) = (\omega_{1},\ldots,\omega_{6})$
\cite{GC},\cite{Van88}. Due to the complementarity of $Sp(12,R)$ and
$O(m)$, the collective states characterized by an $O(m)$ irrep
belong to a single irrep of $Sp(12,R)$. The $Sp(12,R)$ group is
therefore the dynamical group of collective excitations of the
two-component proton-neutron nuclear system. We note that the
reduction $Sp(12m,R) \supset Sp(12,R)\otimes O(m)$ is multiplicity
free
\cite{Filippov81},\cite{Castanos82},\cite{Van88},\cite{Rowe2012}.

The set of basis states of the full dynamical symmetry group
$Sp(12m,R)$ of the whole many-particle nuclear system contains all
possible motions, collective, intrinsic, etc. However, often, one
restricts himself to a certain type of dominating modes in the
process under consideration. Thus, by reducing the group $Sp(12m,R)$
one performs the separation of the nuclear variables into
kinematical (internal) and dynamical (collective) ones. The choice
of the reduction chain depends on the concrete physical problem we
want to consider. As we saw, the group $Sp(12,R)$ plays an important
role in the treatment of the collective excitations of the
proton-neutron nuclear system. The reduction $Sp(12m,R) \supset
Sp(12,R)\otimes O(m)$ turns out to be of a crucial importance in the
microscopic nuclear theory also because the first group in the
direct product subgroup $Sp(12,R)\otimes O(m)$ is associated with
the collective excitations, whereas the second group allows one to
ensure the proper permutational symmetry of the nuclear wave
functions. In this way the considered reduction chain corresponds to
the splitting of the microscopic many-particle configuration space
$\mathbb{R}^{6m}$, spanned by the relative Jacobi vectors, into
kinematical and dynamical submanifolds, respectively.

It is clear then that a large class of collective Hamiltonians
represented by any function of the bilinear combinations of the
coordinates and momenta will lie in the enveloping algebra of the
$Sp(12,R)$ rather than $Sp(12m,R)$. In particular, the kinetic
energy terms for the two subsystems $K(\alpha) =
p^{2}_{\alpha}/2m_{\alpha} = \frac{1}{2m_{\alpha}}T(\alpha,\alpha)
$, their harmonic oscillator Hamiltonians $H_{0}(\alpha) =
\frac{1}{2m_{\alpha}}T(\alpha,\alpha) + \frac{1}{2}
m_{\alpha}\omega^{2}_{\alpha}\sum_{ii}Q_{ii}(\alpha,\alpha)$ are
simply elements of the $Sp(12,R)$ algebra, whereas the collective
potential, represented usually as a function $\upsilon(Q)$ of the
mass quadrupole tensor, will be in the enveloping algebra. With the
microscopic realization of the mass quadrupole (\ref{Sp12a}), the
collective potential $\upsilon(Q)$ becomes a well-defined
shell-model operator. In this way, the $Sp(12,R)$ algebraic
structure embraces both the microscopic collective and the harmonic
oscillator shell-model aspects of the nuclear excitations in the
proton-neutron many-particle systems.

\section{The dynamical content}

We consider the dynamical groups of possible collective flows which
can be generated by the operators (\ref{Sp12a})$-$(\ref{Sp12d}).
They play a fundamental role in the algebraic formulation of the
nuclear collective motion because they are the groups of collective
vibrations and rotations.

In a given collective model, the momentum observables should be the
infinitesimal generators of collective flows corresponding to Lie
group transformations. The operators $L_{ij}(\alpha,\beta)$
(\ref{Sp12c}) are the infinitesimal generators of rigid rotations in
the $6$-dimensional space and are the generators of the group
$SO(6)$. Among them are the $6$ angular momentum components
$L_{k}^{(p)}=\varepsilon_{kij}L_{ij}(p,p)$ and
$L_{k}^{(n)}=\varepsilon_{kij}L_{ij}(n,n)$ ($k,i,j = cyclic$) of the
infinitesimal generators of rigid rotations of the proton and
neutron subsystems, respectively. The remaining $9$ components of
$L_{ij}(\alpha,\beta)$ with $\alpha \neq \beta$ represent combined
proton-neutron collective excitations of the system as a whole.

More general collective flows, e.g. vibrational flows or
irrotational flow rotations, are generated by the infinitesimal
generators of more general dynamical groups. The shear momentum
generators of deformations and rotations $S_{ij}(\alpha,\beta)$
(\ref{Sp12b}) represent the infinitesimal generators of $GL(6,R)$.
The six diagonal momenta of them $S_{ii}(p,p)$ and $S_{ii}(n,n)$ are
infinitesimal generators of monopole and quadrupole shape vibrations
and deformations of the proton and neutron subsystems along the
intrinsic axis $i$, while the off-diagonal components of the shear
momenta $S_{ij}(p,p)$ and $S_{ij}(n,n)$ ($i \neq j$) are
infinitesimal generators of irrotational-flow rotations of the
two-subsystems, respectively. The operators $S_{ii}(p,n)$ and
$S_{ii}(n,p)$ represent a simultaneous deformation of the proton and
neutron distributions (ellipsoids) along the principal axis $i$,
whereas $S_{ij}(p,n)$ and $S_{ij}(n,p)$ generate irrotational-flow
(surface wave) rotations of the combined proton-neutron system.

The operators $S_{ij}(\alpha,\beta)$ together with the angular
momenta $L_{ij}(\alpha,\beta)$ close under commutation and span the
Lie algebra $gl(6,R)$. In this way, the operators
$\{S_{ij}(\alpha,\beta),L_{ij}(\alpha,\beta)\}$ with $\alpha \neq
\beta$ extend the direct sum algebra $gl_{p}(3,R)\oplus
gl_{n}(3,R)$,generated by the set
$\{S_{ij}(\alpha,\alpha),L_{ij}(\alpha,\alpha)\}$ with $\alpha =
p,n$; $i,j = 1,2,3$, to $gl(6,R)$, the algebra of deformations and
rotations in a $6$-dimensional space, including the excitations of
the proton subsystem with respect to the neutron one, as well as
excitations of the combined proton-neutron system as a whole.

If we adjoin the quadrupole moments $Q_{ij}(\alpha,\beta)$ to
$S_{ij}(\alpha,\beta)$ and $L_{ij}(\alpha,\beta)$, we obtain a basis
for the semi-direct sum Lie algebra $gcm(6) =
[\mathbb{R}^{21}]gl(6,R)$, the general collective motion algebra in
six dimensions. The 21 quadrupole moments $Q_{ij}(\alpha,\beta)$
commute among themselves and span the Abelian Lie algebra
$\mathbb{R}^{21}$. They characterize the shape and the orientation
of the proton-neutron nuclear system as a whole in the
six-dimensional space, as well the configuration of the proton
distribution with respect to the neutron one.

The operators $Q_{ij}(\alpha,\beta)$ and $L_{ij}(\alpha,\beta)$
which are subset of $gcm(6)$ generators span the semi-direct sum Lie
algebra $[\mathbb{R}^{21}]so(6)$ which can be denoted as $rot(6)$ by
analogy with $rot(3)=[\mathbb{R}^{5}]so(3)$ algebra. The operators
$Q_{ij}(p,p)$ and $L_{ij}(p,p)$ generate $rot_{p}(3)$ algebra. The
$rot_{n}(3)$ algebra is similarly defined.

Finally, the addition of the $21$ momentum operators
$T_{ij}(\alpha,\beta)$ to the set of $gcm(6)$ algebra generators
extends the latter to the $sp(12,R)$ algebra with total number of
$78$ generators. It is clear that among the generators of $sp(12,R)$
algebra are the proton and neutron kinetic energy operators.

We note that by contraction with respect to $\alpha$ of the
$sp(12,R)$ generators, i.e.
$Q_{ij}=\sum_{\alpha}Q_{ij}(\alpha,\alpha)$,
$S_{ij}=\sum_{\alpha}S_{ij}(\alpha,\alpha)$,
$L_{ij}=\sum_{\alpha}L_{ij}(\alpha,\alpha)$,
$T_{ij}=\sum_{\alpha}T_{ij}(\alpha,\alpha)$, one obtains
respectively the generators of the one-component dynamical algebras
$rot(3)$, $gl(3,R)$, $gcm(3)$, and $sp(6,R)$.

It is clear that a wide class of in-phase (isoscalar) and
out-of-phase (isovector) excitations of the proton subsystem with
respect to the neutron one is present, commonly interpreted in the
IBM-2 \cite{IBM} terms as symmetry and mixed-symmetry states,
respectively. In particular, one has the linear and angular
collective displacements associated with the giant dipole resonance
($E1$ excitations) and the scissors mode ($M1$ excitations) of the
proton system with respect to the neutron one. The latter mode, for
example, is generated by the isovector out-of-phase operator
$\overrightarrow{l} \simeq
(\overrightarrow{L^{p}}-\overrightarrow{L^{n}})$.

The different subgroups of $Sp(12,R)$ reveal some of the possible
classical collective flows. Then it appears that the $Sp(12,R)$
group provides a very general framework in which to investigate the
nature of collective motions in nuclei. To quantize a given
classical collective model one has to construct the irreducible
representations of its dynamical symmetry group. This, as will see,
could be readily done for the $Sp(12,R)$.

Note that all the algebras considered are fully  microscopically
realizable, i.e. they are composed of fully microscopic one-body
operators which act on the Hilbert space of antisymmetrized
many-particle state vectors.

\section{Representations of the $Sp(12,R)$ Lie algebra}

The $Sp(12,R)$ algebra has many nice properties. First, it is a
semi-simple Lie algebra with a well-known representation theory.
Second, important for the shell-model theory of nuclear collective
motion, as we will see, is the fact that its irreps are readily
given in shell-model terms. Indeed, the descrete series
representations are readily constructed on the many-particle Hilbert
state space by the realization of the $Sp(12,R)$ algebra as the
vector space of all skew-adjoint one-body bilinear products in the
position $x_{is}(\alpha)$ and momentum $p_{is}(\alpha)$ observables.
Third, as was shown, the $Sp(12,R)$ algebra contains many of
collective motion algebras as subalgebras which reveal the dynamical
content of the $Sp(12,R)$ model from the hydrodynamic perspective.

To construct the irreducible representations of the $Sp(12,R)$ Lie
algebra in a harmonic oscillator basis, it is convenient to
introduce the harmonic oscillator raising and lowering operators
\begin{eqnarray}
&&b^{\dagger}_{i\alpha,s}=\sqrt{\frac{m_{\alpha}\omega}{2\hbar}}\Big(x_{is}(\alpha)-\frac{i}{m_{\alpha}\omega}p_{is}(\alpha)\Big), \notag\\
&&b_{i\alpha,s}=\sqrt{\frac{m_{\alpha}\omega}{2\hbar}}\Big(x_{is}(\alpha)+\frac{i}{m_{\alpha}\omega}p_{is}(\alpha)\Big),
\label{bos}
\end{eqnarray}
which satisfy the commutation relations
\begin{equation}
[b_{i\alpha,s},b^{\dagger}_{j\beta,t}]=\delta_{ij}\delta_{\alpha\beta}\delta_{st}.
\label{boscom}
\end{equation}
In terms of the harmonic oscillator creation and annihilation
operators, the many-particle realization of the $Sp(12,R)$ Lie
algebra is given by
\begin{eqnarray}
&&F_{ij}(\alpha,\beta)=\sum_{s=1}^{m}b^{\dagger}_{i\alpha,s}b^{\dagger}_{j\beta,s},
\label{F} \\
&&G_{ij}(\alpha,\beta)=\sum_{s=1}^{m}b_{i\alpha,s}b_{j\beta,s},
\label{G} \\
&&A_{ij}(\alpha,\beta)=\frac{1}{2}\sum_{s=1}^{m}(b^{\dagger}_{i\alpha,s}b_{j\beta,s}+b_{j\beta,s}b^{\dagger}_{i\alpha,s}).
\label{A}
\end{eqnarray}
The commutation relation for the $Sp(12,R)$ algebra are easily
inferred from the commutation relations (\ref{boscom}). The
number-conserving operators (\ref{A}) generate the maximal compact
subgroup $U(6)$ of $Sp(12,R)$. We will use also the following
notations $F_{ab} \equiv F_{ij}(\alpha,\beta)$, $G_{ab} \equiv
G_{ij}(\alpha,\beta)$, $A_{ab} \equiv A_{ij}(\alpha,\beta)$ in which
the single indices $a \equiv i\alpha$, $b \equiv j\beta$ are
introduced.

In terms of boson operators (\ref{F})$-$(\ref{A}) the generators
(\ref{Sp12a})-(\ref{Sp12d}) of $Sp(12,R)$ algebra take the form:
\begin{eqnarray}
&&Q_{ij}(\alpha,\beta)=A_{ij}(\alpha,\beta) +
\frac{1}{2}\bigg[F_{ij}(\alpha,\beta) + G_{ij}(\alpha,\beta)\bigg], \label{Sp12a2} \\
&&S_{ij}(\alpha,\beta)=i\bigg[F_{ij}(\alpha,\beta) -
G_{ij}(\alpha,\beta)\bigg], \label{Sp12b2} \\
&&L_{ij}(\alpha,\beta)=-i\bigg[A_{ij}(\alpha,\beta) -
A_{ji}(\beta,\alpha)\bigg], \label{Sp12c2} \\
&&T_{ij}(\alpha,\beta)=A_{ij}(\alpha,\beta) -
\frac{1}{2}\bigg[F_{ij}(\alpha,\beta) + G_{ij}(\alpha,\beta)\bigg].
\label{Sp12d2}
\end{eqnarray}

If we compare the expressions (\ref{F})$-$(\ref{A}) with the
generators (in the coupled angular momentum form) of the
phenomenological algebraic Interacting Vector Boson Model
\cite{IVBM}, it becomes clear that the latter can be considered as
\emph{effective} (or renormalized) counterparts of the microscopic
many-particle operators, defined by (\ref{F})$-$(\ref{A}). However,
an important difference is that by the operators
(\ref{F})$-$(\ref{A}) we can build generic irreducible
representations $E \equiv [E_{1},E_{2},E_{3},E_{4},E_{5},E_{6}]$ of
$U(6)$ in contrast to the IVBM where only the fully symmetric irreps
of $U(6)$ are permitted. As we will see later, this implies some new
features which arise in the present approach. We note also that in
this respect the labeling of the $U(6)$ irreps in the $Sp(12,R)$
scheme proposed here resembles that of IBM-4 \cite{IBM}.

An $Sp(12,R)$ unitary irreducible representation is characterized by
the $U(6)$ quantum numbers  $\sigma=[\sigma_{1},\ldots,\sigma_{6}]$
of its lowest-weight state $|\sigma \rangle$, i.e. $|\sigma \rangle$
satisfies
\begin{eqnarray}
&&G_{ab} |\sigma \rangle = 0, \notag\\
&&A_{ab} |\sigma \rangle = 0, \quad a < b, \notag \\
&&A_{aa} |\sigma \rangle = \sigma_{a} |\sigma \rangle \label{LWS}
\end{eqnarray}
for all $a,b=1,\ldots,6$. Note that the lowest-weight state $|\sigma
\rangle$ for a symplectic irrep is also a highest-weight state for
the $U(6)$ irrep $[\sigma_{1},\ldots,\sigma_{6}]$.

A discrete basis for the irrep $\langle \sigma \rangle \equiv
\langle \sigma_{1}+\frac{n}{2},\ldots,\sigma_{6}+\frac{n}{2}
\rangle$ of $Sp(12,R)$ is generated by the repeated application of
the $Sp(12,R)$ two-boson creation operators on this lowest weight
state. A classification of the states obtained by this procedure is
facilitated by the observation that the raising operators of
$Sp(12,R)$ are components of an inrreducible tensor of $U(6)$.
Indeed, they transform according to the $U(6)$ irreducibe
representation $[2]$. Thus by taking tensor products of these
raising operators, we define the $U(6)$ tensor operators
\begin{equation}
P^{(n)}(F) = [F\times \ldots \times F]^{(n)}, \label{polynom}
\end{equation}
where $n = [n_{1},\ldots,n_{6}]$ is a partition with even integer
parts. It is known that these couplings are multiplicity free. By a
$U(6)$ coupling of these tensor products to the lowest-weight $U(6)$
state $|\sigma \rangle$, one constructs the basis of states for an
$Sp(12,R)$ irrep
\begin{equation}
|\Psi(\sigma n \rho E \eta) \rangle= [P^{(n)}(F) \times |\sigma
\rangle]^{\rho E}_{\eta} , \label{sbasis}
\end{equation}
where $E = [E_{1},\ldots,E_{6}]$ indicates the $U(6)$ quantum
numbers of the coupled state, $\eta$ labels a basis of states for
the coupled $U(6)$ irrep $E$ and $\rho$ is a multiplicity index.
Thus we obtain a basis of $Sp(12,R)$ states that reduce the subgroup
chain $Sp(12,R) \subset U(6)$.

\begin{table}[h]
\caption{The scalar representation $\langle \sigma \rangle = 0$ of
$Sp(12,R)$.} \label{SIR}
\smallskip \centering%
\begin{tabular}{lllll}
\hline &  & $\cdots $ &  &  \\ \hline &  &
$[8],[62],[44],[422],[2222]$ &  &  \\ \hline &  & $[6],[42],[222]$ &
&  \\ \hline &  & $[4],[22]$ &  &  \\ \hline &  & $[2]$ &  &  \\
\hline &  & $[0]$ &  &  \\ \hline
\end{tabular}
\end{table}

As an example, the scalar (i.e. $\langle \sigma \rangle = 0$)
$Sp(12,R)$ irrep is given in Table \ref{SIR}. This example allows us
to compare the representation spaces of the present approach with
that of IVBM \cite{IVBM}. As can be seen, the representation space
of the proton-neutron symplectic model proposed in the present paper
even for the scalar irrep is much richer than that of IVBM, the
latter containing only the fully symmetric $U(6)$ irreps (see e.g.
\cite{GGG}). Generally, for non-scalar irreps of $Sp(12,R)$, some of
the $U(6)$ irreps belonging to the former may appear several times.
Thus a multiplicity index $\rho$ is required, as explicitly shown in
(\ref{sbasis}).

Finally, we note that if we perform a contraction with respect to
the index  $\alpha$, then we obtain the many-particle realization of
the operators of the one-component nuclear system $F_{ij} =
\sum_{\alpha} F_{ij}(\alpha,\alpha)$, $G_{ij} = \sum_{\alpha}
G_{ij}(\alpha,\alpha)$ and $A_{ij} = \sum_{\alpha}
A_{ij}(\alpha,\alpha)$, which generate the group $Sp(6,R)$. In other
words, one obtains the $Sp(6,R)$ model \cite{RR1} as a submodel of
the $Sp(12,R)$ one, in contrast to the microscopic $Sp(12,R)$ model
introduced in Ref.\cite{Van75}, in which the components of the mass
quadrupole tensor are used as collective variables. Expressing the
latter and their derivatives through the boson creation and
annihilation operators, among the reduction chains considered in
\cite{Van75}, the three algebraic structures ($U(5)$, $O(6)$ and
$SU(3)$) of the IBM-1 \cite{IBM} were obtained, which are embedded
in $Sp(12,R)$ through the group $U(6) \subset Sp(12,R)$ associated
with the 6 quadrupole collective degrees of freedom.

\section{The shell-model classification of nuclear collective states}

The relevant $Sp(12,R)$ irreducible representations appropriate for
the description of the low-lying collective states in heavy mass
deformed nuclei - and correspondingly the related $O(m)$ irreps - in
the reduction (\ref{SpO}) can be fixed by considering the underlying
shell-model structure of the ground state. To reveal this structure
we specify the basis of an $Sp(12,R)$ irrep by considering the
following reduction of the subgroup $U(6) \subset Sp(12,R)$:
\begin{eqnarray}
Sp(12,R) && \supset \notag \\
\sigma \quad &&  n\rho \notag \\
\notag \\
&&\supset U(6) \supset \ SU_{p}(3) \otimes SU_{n}(3) \notag \\
&&\quad\quad E \ \ \gamma \quad (\lambda_{p},\mu_{p}) \ \
(\lambda_{n},\mu_{n}) \ \notag \\
\notag \\
&&\supset SU(3) \supset SO(3) \supset SO(2). \notag \\
&&\varrho(\lambda,\mu) \ \ K \quad L \quad\quad\quad M
\label{SU3xSU3}
\end{eqnarray}
The chain (\ref{SU3xSU3}) naturally generalizes the Elliott's
$SU(3)$ model \cite{Elliott58} by extending the model space to the
direct product space $SU_{p}(3) \otimes SU_{n}(3)$ of proton and
neutron subsystems. The $SU(3)$ irreps of the two subsystems are
subsequently coupled to the $SU(3)$ irrep of the combined
proton-neutron system. The combined $SU(3)$ algebra is generated by
the quadrupole $Q_{M}=Q^{p}_{M}+Q^{n}_{M}$ and angular momentum
$L_{M}=L^{p}_{M}+L^{n}_{M}$ operators, respectively. The chain
(\ref{SU3xSU3}) corresponds to the following choice of the index
$\eta=\gamma(\lambda_{p},\mu_{p})(\lambda_{n},\mu_{n})\varrho(\lambda,\mu)KLM$
labeling the basis states (\ref{sbasis}) of an $Sp(12,R)$ irrep.

The choice of the coupling scheme (\ref{SU3xSU3}) is dictated by the
fact that the dominant component of the nuclear interaction in heavy
mass deformed nuclei is provided by the quadrupole-quadrupole
forces. The Eq.(\ref{SU3xSU3}) implies a strong coupling of the
proton and neutron distributions to form a composite distribution of
the combined proton-neutron system with different possible
deformations. The maximum deformation is obtained by restricting the
direct product irrep $(\lambda_{p},\mu_{p}) \otimes
(\lambda_{n},\mu_{n})$ of $SU_{p}(3) \otimes SU_{n}(3)$ to the
leading irreducible representation
$(\lambda_{p}+\lambda_{n},\mu_{p}+\mu_{n})$ of $SU(3)$. Then the
corresponding geometric picture of the algebraic structure defined
by (\ref{SU3xSU3}) is that of two coupled rotors (one rotor
representing the protons and another for the neutrons)
\cite{TRM},\cite{Castanos87}. We stress that the reduction of a
generic $U(6)$ irreducible representation $E \equiv
[E_{1},E_{2},E_{3},E_{4},E_{5},E_{6}]$ to the direct product irreps
of $SU_{p}(3) \otimes SU_{n}(3)$ allows irreps of the type
$(\lambda_{p},\mu_{p}) \otimes (\lambda_{n},\mu_{n})$ with nonzero
values of the quantum numbers $\lambda$ and $\mu$ characterizing the
proton and neutron $SU(3)$ irreps. The latter geometrically
corresponds to two non-axial rotors \cite{Castanos87}. This is in
contrast to the case of the IVBM \cite{IVBM},\cite{GGG} in which the
model space is spanned only by all fully symmetric $U(6)$ irreps
that reduce to the $SU_{p}(3) \otimes SU_{n}(3)$ direct product
irreps of the type $(\lambda_{p},0) \otimes (\lambda_{n},0)$. Thus,
the geometric picture of the latter is that of two coupled axial
rotors \cite{Castanos87}.

The generators of $Sp(12,R)$ (\ref{F})$-$(\ref{A}) can be classified
as irreducible tensor operators with respect to different subgroups
of the whole chain (\ref{SU3xSU3}) and hence will be characterized
by the quantum numbers determining their irreducible
representations. For the raising operators one readily obtains the
following tensors:
\begin{eqnarray}
&&F^{[2]_{6}\quad\quad\quad LM}_{(2,0)(0,0) \quad (2,0)}(p,p), \quad
F^{[2]_{6}\quad\quad\quad LM}_{(0,0)(2,0) \quad (2,0)}(n,n), \notag \\
\notag \\
&&F^{[2]_{6}\quad\quad\quad LM}_{(1,0)(1,0) \quad (2,0)}(p,n),
\label{Ft1}
\end{eqnarray}
where $L=0,2; M=-L,\ldots,M$ , and
\begin{equation}
F^{[2]_{6}\quad\quad\quad 1M}_{(1,0)(1,0) \quad (0,1)}(p,n).
\label{Ft0}
\end{equation}
We see a multiplication of the standard one-component $Sp(6,R)$
raising generators \cite{RR1} which for the two-component system
correspond to the creation of monopole and quadrupole $pp$, $nn$,
and $pn$ pairs. In addition to the $(2,0)$ $SU(3)$ raising
generators $F^{[2]_{6}\quad\quad\quad LM}_{(p,0)(q,0) \quad
(2,0)}(\alpha,\beta)$ (\ref{Ft1}) we have also the $(0,1)$ $SU(3)$
tensor operator $F^{[2]_{6}\quad\quad\quad 1M}_{(1,0)(1,0) \quad
(0,1)}(p,n)$ (\ref{Ft0}), which is a new one compared to the
generators of the $Sp(6,R)$ model of Rosensteel and Rowe \cite{RR1}.

The number of bosons operator $N$ is the first Casismir invariant of
the $U(6)$ as well as of the combined proton-neutron $U(3)$ group.
The latter allows us to determine the shell-model tensor properties
by considering the reduction chain $U(3) \supset U(1) \otimes
SU(3)$. Thus, in shell-model terms, the raising operators of
$Sp(12,R)$ with $U(1) \otimes SU(3)$ quantum numbers $N(\lambda,\mu)
= 2(2,0)$ and their conjugate lowering ones represent $\pm
2\hbar\omega$ inter-shell collective excitations of monopole and
quadrupole type. Additionally, in contrast to the $Sp(6,R)$ model,
the $Sp(12,R)$ raising operators with $N(\lambda,\mu) = 2(0,1)$
together with their conjugate lowering operators correspond to the
$\pm 2\hbar\omega$ inter-shell excitations of dipole type. Thus, the
$Sp(12,R)$ collective dynamics covers the nuclear coherent
excitations of monopole, dipole and quadrupole type.

The basis states classified according to (\ref{SU3xSU3}) can be
written as
\begin{equation}
| N_{min};\sigma n\rho E; \gamma
(\lambda_{p},\mu_{p})(\lambda_{n},\mu_{n})\varrho(\lambda,\mu); KLM
\rangle, \label{bas}
\end{equation}
where $\rho$, $\gamma$ and $\varrho$ are multiplicity indices.
Recall that $\sigma = [\sigma_{1},\ldots,\sigma_{6}]$, $n =
[n_{1},\ldots,n_{6}]$, $E = [E_{1},\ldots,E_{6}]$. These basis
states can be simultaneously classified according to the chain
(\ref{SpO}). Then the symplectic bandhead structure determined by
the $U(6)$ irrep $\sigma$ will coincide with the $O(m)$ irrep
$\omega$, i.e. $\sigma \equiv \omega$ \cite{Van88}.

The  appearance of the group $O(m)$ in (\ref{SpO}) turns out to be
crucial because it allows one to construct the nuclear wave
functions with the proper permutational symmetry. The essential
property that makes it possible is the fact that the group $O(m)$
contains the symmetric group $S_{m+1}$ as a subgroup. Thus, to fix
the permutational symmetry of the wave function, we consider the
embedding of the symmetric group $S_{m+1}$ in the $O(m)$ according
to
\begin{align}
&O(m) \supset S_{m+1}, \notag \\
&\quad \omega \quad \delta \quad [f]h \label{Sn}
\end{align}
where $[f]$ is the Young scheme characterizing the irreducible
representation of the permutational group $S_{m+1}$, $h$ indexes its
basis, and $\delta$ is a multiplicity index. However, because the
antisymmetry should be satisfied separately for protons and
neutrons, in order to insure the proper permutational symmetry we
consider further the reduction of $S_{A}$ to $S_{N_{1}} \otimes
S_{N_{2}}$ ($A = N_{1}+N_{2}$), i.e.
\begin{equation}
S_{A} \supset S_{N_{1}} \otimes S_{N_{2}}. \label{Sn1xSn2}
\end{equation}
Taking this into account we replace $h$ by
$\delta_{0}[f_{1}]h_{1}[f_{2}]h_{2}$, where $\delta_{0}$ is a
multiplicity index in the reduction (\ref{Sn1xSn2}). The full
antisymmetry of the total wave function is therefore ensured by
coupling of the Young scheme of the $S_{A}$ irrep to its conjugate
(contragradient) representation of the spin wave function.

The basis states classified according to (\ref{SpO}),
(\ref{SU3xSU3}) and (\ref{Sn}) can  be written as
\begin{equation}
| N_{min};n\rho E;\gamma
(\lambda_{p},\mu_{p})(\lambda_{n},\mu_{n})\varrho(\lambda,\mu); KLM;
\omega\delta[f]h \rangle, \label{bas2}
\end{equation}
where $h$ is a basis of the $S_{m+1}$ irrep $[f]$, which is further
fixed by the reduction chain (\ref{Sn1xSn2}). In (\ref{bas}) and
(\ref{bas2}), $N_{min}$ counts the minimum number of oscillator
quanta (phonons) allowed by the Pauli principle.

As we saw, a generic $Sp(12,R)$ irrep is determined by the $U(6)$
lowest weight with $\sigma = [\sigma_{1},\ldots,\sigma_{6}]$ and
contains all $U(6)$ irreps $E = [E_{1},\ldots,E_{6}]$ which are
obtained by the $U(6)$-coupling $[\sigma_{1},\ldots,\sigma_{6}]
\otimes [n_{1},\ldots,n_{6}]$. However, one expects the most
symmetric $U(6)$ irreps $E$ represented by the one- and two-rowed
Young schemes to be dominant in the low-energy spectra of the heavy
deformed even-even nuclei. As an example, the symplectic
classification of the $SU(3)$ basis states according to the
decompositions given by the chain (\ref{SU3xSU3}) for the scalar
$Sp(12,R)$ irrep $\langle \sigma \rangle = 0$, restricted to the
two-rowed $U(6)$ partitions is given in Table \ref{basis}. We see
that even for the scalar $Sp(12,R)$ representation one has a very
rich algebraic structure of the state space. For comparison, the
corresponding $SU(3)$ basis states for the number of oscillator
quanta $N = 0, 2, 4, \ldots$ contained in the scalar irreducible
representation of the $Sp(6,R)$ model \cite{RR1} are marked in red.

Since the collective states of the $Sp(12,R)$ irreducible spaces for
heavy deformed nuclei are constructed from the excitations built on
the two adjacent major shell intrinsic structures with opposite
parity, then the collective spaces obviously consists of both the
positive and negative parity excitations. From Table \ref{basis} we
see the appearance of many new $SU(3)$ multiplets which contain a
richer angular momentum and parity content, as well as a
multiplication of the $SU(3)$ irreps arising from the coupling of
different initial proton and neutron configurations and hence giving
rise to distinct coupled proton-neutron $SU(3)$ collective
excitations with both the positive and negative parity.

In geometrical terms, from Table \ref{basis} we see that except the
two-axial and axial-triaxial, the two-triaxial rotor model
configurations, mentioned above, also appear already in the
decomposition of the two-rowed $U(6)$ irreps for oscillator quanta
$N \geq 6$. It is clear that the $U(6)$ Young schemes with more than
two rows will also give rise to the two-triaxial rotor model
configurations.

In general, a nonscalar $Sp(12,R)$ irreducible representation
$\langle \sigma \rangle \neq 0$ will corresponds to a given real
nucleus. This is a very important feature of the present $Sp(12,R)$
collective model, which is a consequence of the two-component
composite character of the nuclear systems. A given $Sp(12,R)$
irreducible representation, as was discussed, is determined by the
corresponding symplectic bandhead (or intrinsic) structure defined
by the lowest $U(6)$ irrep $\sigma = [\sigma_{1},\ldots,\sigma_{6}]
\neq 0$. The latter, in contrast to the $Sp(6,R)$ case, will contain
a plethora of $SU(3)$ multiplets. This is of significant importance
in the microscopic nuclear structure theory because the states
belonging to the intrinsic space spanned by the $U(6)$ irrep
$[\sigma_{1},\ldots,\sigma_{6}]$ will contain all the necessary
$SU(3)$ irreducible representations needed for the description of
different low-lying collective bands (ground state, $\beta$,
$\gamma$, $K^{\pi} = 0^{-}$, $1^{-}$, $2^{-}$, etc.) in the spectra
of heavy even-even deformed nuclei. In this way, in contrast to the
$Sp(6,R)$ model, the intrinsic $Sp(12,R)$ bandhead structure
provides us with a framework for the simultaneous shell-model
interpretation of the ground state band and the other excited
low-lying collective bands without the need of involving the mixing
of different symplectic irreps (c.f. Ref.\cite{Carvalho86}).

How the intrinsic $U(6)$ structure can be determined in practice for
a certain nucleus? The proper choice is suggested by the shell
model. In this way, for a given nucleus the appropriate symplectic
bandhead can be determined by fixing the corresponding underlying
proton-neutron shell-model structure $SU_{p}(3) \otimes SU_{n}(3)
\supset SU(3)$ embedded in the $U(6)$ irrep
$[\sigma_{1},\ldots,\sigma_{6}]$. The parent $SU(3)$ irreps
$(\lambda_{p},\mu_{p})$ and $(\lambda_{n},\mu_{n})$ of the two
subsystems, which are consequently strongly coupled to the $SU(3)$
irrep $(\lambda,\mu)$ of the combined proton-neutron nuclear system
as a whole, are determined by compactly filling pairwise the
$3$-dimensional harmonic oscillator potential with protons and
neutrons, respectively. Then $N_{min}$ in (\ref{bas}) and
(\ref{bas2}) will counts the total number of oscillator quanta
consistent with the Pauli principle, counting all filled levels and
remembering to include the factor $\frac{6}{2}m$ for the zero-point
motion of the $m = A-1$ quasiparticles associated with the relative
Jacobi vectors, i.e. $N_{min} = (\sigma_{1} + \ldots + \sigma_{6}) +
\frac{6}{2}m$ \cite{Filippov81},\cite{Nmin1}-\cite{Nmin4}.

\begin{table}[h]
\caption{Symplectic classification of the $SU(3)$ basis states.}
\centering \label{basis}
\smallskip\centering\small\addtolength{\tabcolsep}{-0pt}
\begin{tabular}{||l||l||l|l|l||}
\hline\hline $N$ & $[E_{1},\ldots ,E_{6}]$ & $(\lambda _{p},\mu
_{p})$ & $(\lambda
_{n},\mu _{n})$ & $\ \ \ \ \ \ \ \ \ \ \ \ \ \ \ \ \ \ (\lambda ,\mu )$ \\
\hline\hline $0$ & $[0]$ & $\ (0,0)$ & $\ (0,0)$ & $\
{\color{red}(0,0)}$ \\ \hline $2$ & $[2]$ &
\begin{tabular}{l}
$(2,0)$ \\
$(1,0)$ \\
$(0,0)$%
\end{tabular}
&
\begin{tabular}{l}
$(0,0)$ \\
$(1,0)$ \\
$(2,0)$%
\end{tabular}
&
\begin{tabular}{l}
$(2,0)$ \\
${\color{red}(2,0)},(0,1)$ \\
$(2,0)$%
\end{tabular}
\\ \hline
& $[4]$ &
\begin{tabular}{l}
$(4,0)$ \\
$(3,0)$ \\
$(2,0)$ \\
$(1,0)$ \\
$(0,0)$%
\end{tabular}
&
\begin{tabular}{l}
$(0,0)$ \\
$(1,0)$ \\
$(2,0)$ \\
$(3,0)$ \\
$(4,0)$%
\end{tabular}
&
\begin{tabular}{l}
$(4,0)$ \\
$(4,0),(2,1)$ \\
${\color{red}(4,0)},(2,1),{\color{red}(0,2)}$ \\
$(4,0),(2,1)$ \\
$(4,0)$%
\end{tabular}
\\ \cline{2-5}
$%
\begin{tabular}{l}
$4$ \\
\\
\\
\\
\\
\end{tabular}%
$ & $[22]$ &
\begin{tabular}{l}
$(0,1)$ \\
$(1,1)$ \\
$(1,0)$ \\
$(0,2)$ \\
$(0,0)$ \\
$(2,0)$%
\end{tabular}
&
\begin{tabular}{l}
$(0,1)$ \\
$(1,0)$ \\
$(1,1)$ \\
$(0,0)$ \\
$(0,2)$ \\
$(2,0)$%
\end{tabular}
&
\begin{tabular}{l}
$(0,2),(1,0)$ \\
$(2,1),(0,2),(1,0)$ \\
$(2,1),(0,2),(1,0)$ \\
$(0,2)$ \\
$(0,2)$ \\
$(4,0),(2,1),(0,2)$%
\end{tabular}
\\ \hline
& $[6]$ &
\begin{tabular}{l}
$(6,0)$ \\
$(5,0)$ \\
$(4,0)$ \\
$(3,0)$ \\
$(2,0)$ \\
$(1,0)$ \\
$(0,0)$%
\end{tabular}
&
\begin{tabular}{l}
$(0,0)$ \\
$(1,0)$ \\
$(2,0)$ \\
$(3,0)$ \\
$(4,0)$ \\
$(5,0)$ \\
$(6,0)$%
\end{tabular}
&
\begin{tabular}{l}
$(6,0)$ \\
$(6,0),(4,1)$ \\
$(6,0),(4,1),(2,2)$ \\
${\color{red}(6,0)},(4,1),{\color{red}(2,2)},(0,3)$ \\
$(6,0),(4,1),(2,2)$ \\
$(6,0),(4,1)$ \\
$(6,0)$%
\end{tabular}
\\ \cline{2-5}
$6$ & $[42]$ &
\begin{tabular}{l}
$(2,2)$ \\
$(1,2)$ \\
$(0,2)$ \\
$(2,1)$ \\
$(1,1)$ \\
$(0,1)$ \\
$(3,1)$ \\
$(2,1)$ \\
$(1,1)$ \\
$(2,0)$ \\
$(1,0)$ \\
$(0,0)$ \\
$(3,0)$ \\
$(2,0)$ \\
$(1,0)$ \\
$(4,0)$ \\
$(3,0)$ \\
$(2,0)$%
\end{tabular}
&
\begin{tabular}{l}
$(0,0)$ \\
$(1,0)$ \\
$(2,0)$ \\
$(0,1)$ \\
$(1,1)$ \\
$(2,1)$ \\
$(1,0)$ \\
$(2,0)$ \\
$(3,0)$ \\
$(0,2)$ \\
$(1,2)$ \\
$(2,2)$ \\
$(1,1)$ \\
$(2,1)$ \\
$(3,1)$ \\
$(2,0)$ \\
$(3,0)$ \\
$(4,0)$%
\end{tabular}
&
\begin{tabular}{l}
$(2,2)$ \\
$(2,2),(0,3),(1,1),(0,0)$ \\
$(2,2),(1,1),(0,0)$ \\
$(2,2),(0,3),(1,1),(0,0)$ \\
$(2,2),2(1,1),(0,0),(3,0),(0,3)$ \\
$(2,2),(3,0),(1,1),(0,0)$ \\
$(4,1),(2,2),(0,3),(3,0),(0,1)$ \\
$(4,1),(2,2),(0,3),(3,0),(0,1)$ \\
$(4,1),(2,2),(0,3),(3,0),(0,1)$ \\
$(2,2),(1,1),{\color{red}(0,0)}$ \\
$(2,2),(0,3),(1,1),(0,0)$ \\
$(2,2)$ \\
$(4,1),(2,2),(0,3),(3,0),(0,1)$ \\
$(4,1),(2,2),(0,3),(3,0),(0,1)$ \\
$(4,1),(2,2),(0,3),(3,0),(0,1)$ \\
$(6,0),(4,1),(2,2)$ \\
$(6,0),(4,1),(2,2),(0,3)$ \\
$(6,0),(4,1),(2,2)$%
\end{tabular}
\\ \hline
\ldots  & \ldots  & \ldots  & \ldots  & \ldots  \\ \hline\hline
\end{tabular}
\end{table}

Having the basis, an arbitrary microscopic or phenomenological
Hamiltonian can be diagonalized within the collective symplectic
space of $Sp(12,R)$ algebra. The calculation is simplified when the
Hamiltonian lies in the enveloping algebra of $Sp(12,R)$ since the
latter contains many physically relevant parts of the nuclear
forces, like the proton and neutron harmonic oscillator
Hamiltonians, the kinetic energy terms for the two subsystems, the
collective potential represented by a scalar function of the full
quadrupole operator, and a residual interaction. The latter should
include, for example, single-particle spin-orbit and orbit-orbit
terms, as well as pairing and other interactions. In practical
calculations, however, the Hamiltonian can be restricted to form
that is solely expressed in terms of the symplectic generators.
Interaction of this form do not mix different symplectic irreps and
the Hamiltonian for such interactions will have block-diagonal
structure. The single symplectic irrep approximation will be a
sensible choice for nuclear systems that have a dominant
quadrupole-quadrupole force. The latter favors the states with
maximum spatial symmetry and the largest value of the second
invariant of the $SU(3)$.

Following the concepts of Refs.\cite{Carvalho86},\cite{Rowe85} we
can define the \emph{collective subspaces} (vertical cones) as the
irreducible symplectic subspaces of the nuclear Hilbert space. Each
$Sp(12,R)$ irrep is characterized by a lowest-weight state with
quantum numbers $\sigma = [\sigma_{1},\ldots,\sigma_{6}]$. Then if
$\{|\sigma\eta \rangle\}$ denotes a basis for the $U(6)$
lowest-weight space of an $Sp(12,R)$ irrep $\sigma$, any shell-model
state belonging to this collective subspace can be expressed as
\begin{equation}
\psi_{\sigma} = \sum_{\eta} \psi_{\eta}(F)|\sigma\eta \rangle,
\label{sms}
\end{equation}
where $\psi_{\eta}$ is a polynomial in the $Sp(12,R)$ raising
operators. The lowest-weight state of an $Sp(12,R)$ irrep is
referred to as an intrinsic state for that collective subspace. The
extension to an arbitrary shell-model state expresses the fact that
the shell-model space can be decomposed into a direct sum of
$Sp(12,R)$ irreps. Correspondingly, the $Sp(12,R)$ \emph{symplectic
shells} (horizontal layers) are defined as the direct sum of all $n
\hbar\omega$ states of fixed $n$, which are obtained by the repeated
action of the $Sp(12,R)$ raising operators on the $U(6)$ intrinsic
space states. In this way the nuclear Hilbert space naturally
divides simultaneously into vertical cones and horizontal layers,
reflecting the collective and single-particle aspects of nuclear
motion.

We note that the Eq.(\ref{sms}) can be interpreted as a factoring of
an arbitrary wave function into collective and intrinsic parts. The
states $|\sigma\eta \rangle$ can be thought of as intrinsic states
and the rasing operators $\psi_{\eta}$ as collective wave functions.
If the Hamiltonian under consideration consists of terms that mix
different symplectic irreps (horizontal mixing), then a sum over
$\sigma$ in the Eq.(\ref{sms}) should be included.

Concluding, we want to point out that other possibilities exist to
arrange the low-lying symplectic irreps in the low-lying energy
spectra and to fix the intrinsic structure of the ground state. The
relevant symplectic bandhead intrinsic structure can be determined
by taking into account the proper deformation using the deformed
harmonic oscillator (asymptotic Nilsson model \cite{Nilsson}) or
pseudo-$SU(3)$ \cite{pseudo-SU3} schemes of filling the proton and
neutron single particle levels.

Finally, in practical calculations, one may use other reductions of
the $U(6)$ group (e.g., through the $O(6)$ group appropriate for
transitional, $\gamma$-unstable, nuclei) to classify the basis
states, also consistent with the underlying proton-neutron
shell-model structure.

\section{The spin part}

The internal degrees of freedom associated with the group $O(m)$
play an important role in the construction of the microscopic wave
functions because they allow one to ensure the full antisymmetry of
the total wave function. This is achieved by coupling of the Young
scheme of the $S_{m}$ irrep to its conjugate (contragradient)
representation of the spin wave-function.

The construction of the spin function is reduced to the coupling of
the two subsystems spins $S_{1}$ and $S_{2}$ into total spin $S$ of
the nucleus \cite{Van88},\cite{Van71}:
\begin{align}
&\Phi\Big(
[\widetilde{f}]\widetilde{\delta_{0}}[\widetilde{f_{1}}]\widetilde{h_{1}}[\widetilde{f_{2}}]\widetilde{h_{2}};SM_{S}\Big) \notag \\
& =
\sum_{M_{S_{1}}M_{S_{2}}}\Phi\Big([\widetilde{f_{1}}]\widetilde{h_{1}};S_{1}M_{S_{1}}\Big)
 \Phi\Big([\widetilde{f_{2}}]\widetilde{h_{2}};S_{2}M_{S_{2}}\Big) \notag \\
&\quad\quad \times C^{S_{1} \quad S_{2} \quad
 S}_{M_{S_{1}} \ M_{S_{2}} \ M_{S}}. \label{SF}
\end{align}
The Pauli principle requires the antisymmetry of the wave function
with respect to proton and neutron variables separately. That is why
only the proton and neutron schemes $[f_{1}]$,$[\widetilde{f_{1}}]$
and $[f_{2}]$,$[\widetilde{f_{2}}]$ are coupled separately to the
antisymmetric irreps $a_{1}$ and $a_{2}$, respectively. Then the
full antisymmetric wave function can be written in the following
form
\begin{align}
&|N_{min};n\rho E;\gamma
(\lambda_{p},\mu_{p})(\lambda_{n},\mu_{n})\varrho(\lambda,\mu); K(LS)JM_{J}; \notag \\
\notag \\
&\omega\delta[f]\delta_{0}[f_{1}][f_{2}] \rangle \notag \\
\notag \\
&= \sum_{h_{1}h_{2}MM_{S}}|N_{min};n\rho E;\gamma
(\lambda_{p},\mu_{p})(\lambda_{n},\mu_{n})\varrho(\lambda,\mu); \notag \\
&KLM; \omega\delta[f]\delta_{0}[f_{1}]h_{1}[f_{2}]h_{2} \rangle  \notag \\
&\times \Phi\Big(
[\widetilde{f}]\widetilde{\delta_{0}}[\widetilde{f_{1}}]\widetilde{h_{1}}[\widetilde{f_{2}}]\widetilde{h_{2}};SM_{S}\Big)
C^{[f_{1}] \ [\widetilde{f_{1}}] \ a_{1}}_{h_{1} \quad
\widetilde{h_{1}} \ h_{a_{1}}} C^{[f_{2}] \ [\widetilde{f_{2}}] \
a_{2}}_{h_{2} \quad \widetilde{h_{2}} \ h_{a_{2}}} \notag \\
&\times C^{[f]\delta_{0} \ [f_{1}] \ [f_{2}]}_{\ h \quad \ \ h_{1} \
\ h_{2}} C^{L \quad S \quad J}_{M \ \ M_{S} \ M_{J}}, \label{TASF}
\end{align}
where $C^{[f_{1}] \ [\widetilde{f_{1}}] \ a_{1}}_{h_{1} \quad
\widetilde{h_{1}} \ h_{a_{1}}}$, $C^{[f_{2}] \ [\widetilde{f_{2}}] \
a_{2}}_{h_{2} \quad \widetilde{h_{2}} \ h_{a_{2}}}$, $C^{L \quad S
\quad J}_{M \ \ M_{S} \ M_{J}}$ are respectively the Clebsch-Gordan
coefficients for the groups $S_{N_{1}}$, $S_{N_{2}}$ and $SU(2)$,
and $C^{[f]\delta_{0} \ [f_{1}] \ [f_{2}]}_{\ h \quad \ \ h_{1} \ \
h_{2}}$ is the isoscalar factor for the chain $S_{A} \supset
S_{N_{1}} \otimes S_{N_{2}}$.

With this, the task of constructing the fully microscopic
antisymmetric wave functions of the proton-neutron $Sp(12,R)$ model
of nuclear collective motions is completed. The latter allows the
spin contribution of different parts of the nuclear interaction
(e.g. spin-orbit, vector, etc. forces) to be included in the
consideration, as well as to encompass the treatment of the odd-mass
and odd-odd nuclei together with the even-even ones within a single
framework.

\section{Conclusions}

In the present paper, a proton-neutron symplectic model of
collective motions, based on the non-compact symplectic group
$Sp(12,R)$, is introduced by considering the symplectic geometry of
the two-component many-particle nuclear system. The non-compact
feature of symplectic scheme is an essential ingredient of the model
that allows the theory to accommodate quadrupole coherence which
develop in the collective dynamics.

The problem of $6m$ degrees of freedom defined in the many-particle
Hilbert space can be associated with a definite irrep of the
dynamical symmetry group $Sp(12m,R)$. It was proved, however, that
the collective effects are associated with operators that are scalar
in $O(m)$. Then, the collective part of the Hamiltonian can be
obtained by projecting the latter on a definite $O(m)$ irrep
associated with the $m$ Jacobi vectors in the configuration space
$\mathbb{R}^{6m}$, where $m=A-1$ and $A$ is equal to the total
number of nucleons in the system.

From the hydrodynamic perspective, the possible classical collective
motions are determined by different dynamical groups that can be
constructed from the symplectic generators of $Sp(12,R)$, including
a wide class of both the in-phase (isoscalar) and out-of-phase
(isovector) excitations of the proton subsystem with respect to the
neutron one, as well as collective excitations of the combined
proton-neutron system as a whole. The $Sp(12,R)$ group provides
therefore a general framework for the investigation of the nature of
classical collective motions in nuclei.

The relation of the $Sp(12,R)$ irreps with the shell-model
classification of the basis states is considered by extending of the
model space to the direct product space of $SU_{p}(3) \otimes
SU_{n}(3)$ irreps, generalizing in this way the Elliott's $SU(3)$
model for the case of two-component system. The $Sp(12,R)$ model
appears then as a natural multi-major-shell extension of the
generalized proton-neutron $SU(3)$ scheme which takes into account
the core collective excitations of monopole, quadrupole and dipole
type associated with the giant vibrational degrees of freedom. Each
$Sp(12,R)$ irreducible representation is determined by a symplectic
bandhead or an intrinsic $U(6)$ space which can be fixed by the
underlying proton-neutron shell-model structure, so the theory
becomes completely compatible with the Pauli principle. It is shown
that this intrinsic $U(6)$ structure is of vital importance for the
appearance of the low-lying collective bands with both the positive
and negative parity. The full range of low-lying collective states
can then be described by the microscopically based intrinsic $U(6)$
structure, renormalized by coupling to the giant resonance
vibrations.

Summarizing, the $Sp(12,R)$ symplectic model provides a natural
framework for the simultaneous macroscopic and microscopic
description of nuclear collective dynamics of the two-component
proton-neutron nuclear systems.


\begin{thebibliography}{99}

\bibitem{DG} \emph{Dynamical Groups and Spectrum generating Algebras}, Volumes 1 and 2,
edited by A. Bohm, Y. Ne'eman, and A. O. Barut (World Scientific
Publishing Co. Pte. Ltd., Singapore, 1988).

\bibitem{aans} \emph{Algebraic Approaches to Nuclear Structure:
Interacting Boson and Fermion Models}, edited by R. F. Casten
(Harwood Academic Publishers, New York, 1993).

\bibitem{Iachello} F. Iachello, \emph{Lie Algebras and Applications}, Lect.
Notes Phys. Volume 708 (Springer-Verlag, Berlin Heidelberg, 2006).

\bibitem{Frank} \emph{Symmetries in Atomic Nuclei: From Isospin to
Supersymmetry}, by A. Frank, P. V. Isacker, J. Jolie, Springer
Tracks in Modern Physics Volume \textbf{230} (Springer-Verlag, New
York, 2009).

\bibitem{RW} \emph{Fundamentals of Nuclear Models: Foundational Models},
by D. J. Rowe, J. L. Wood (World Scientific Publisher Press,
Singapore, 2010).

\bibitem{NA1} G. Rosensteel and D. J. Rowe, Ann. Phys. \textbf{96}, 1
(1976).

\bibitem{NA2} G. Rosensteel and D. J. Rowe, Ann. Phys. \textbf{123},
36 (1978).

\bibitem{orbits}  D. J. Rowe and G. Rosensteel,
Ann. Phys. \textbf{126}, 198 (1980).

\bibitem{NA4} O. L. Weaver, R. Y. Cussion and L. C. Biedenharn, Ann. Phys. (N.Y.) \textbf{102},
493 (1976).

\bibitem{sl3R} L. Weaver and L. C. Biedenharn,
Nucl. Phys. \textbf{A185}, 1 (1972).

\bibitem{rot3} H. Ui, Prog. Theor. Phys. \textbf{44}, 153 (1970).

\bibitem{cm3} L. Weaver, R. Y. Cusson and L. C. Biedenharn, Ann.
Phys. (N.Y.) \textbf{77}, 250 (1973).

\bibitem{RR1}  D. J. Rowe and G. Rosensteel,
Phys. Rev. Lett. \textbf{38}, 10 (1977).

\bibitem{Van74} V. V. Vanagas,
\emph{Methods of the theory of group representations and separation
of collective degrees of freedom in the nucleus}, Lecture notes at
the Moscow Engineering Physics Institute (MIFI, Moscow, 1974) (in
russian).

\bibitem{Van76} V. Vanagas, Fiz. Elem. Chastits At. Yadra. \textbf{7}, 309
(1976) [Sov. J. Part. Nucl. \textbf{7}, 118 (1976)].

\bibitem{Elliott58} J. P. Elliott,
Proc. R. Soc. London, Ser. A \textbf{245}, 128 (1958); \textbf{245},
562 (1958).

\bibitem{Van80} V. Vanagas, Fiz. Elem. Chastits At. Yadra. \textbf{11},
454 (1976) [Sov. J. Part. Nucl. \textbf{11},  (1980)].

\bibitem{Filippov81} G. F. Filippov, V. I. Ovcharenko, and Yu. F.
Smirnov, \emph{Microscopic Theory of Collective Excitations in
Nuclei} (Naukova Dumka, Kiev, 1981) (in Russian).

\bibitem{Castanos82} O. Castanos, A. Frank, E. Chacon, P. Hess, and
M . Moshinsky, J. Math. Phys. \textbf{23}, 2537 (1982).

\bibitem{GC} M. Moshinsky, J. Math. Phys. \textbf{25}, 1555 (1984).

\bibitem{MQ}  M. Moshinsky and C. Quesne,
J. Math. Phys. \textbf{11}, 1631 (1970).

\bibitem{Van88} V. V. Vanagas,
\emph{Algebraic foundations of microscopic nuclear theory} (Nauka,
Moscow, 1988) (in russian).

\bibitem{Rowe2012} D. J. Rowe, M. J. Carvalho, and J. Repka, Rev.
Mod. Phys. \textbf{84}, 711 (2012).

\bibitem{IVBM} A. Georgieva, P. Raychev, and R. Roussev,
J. Phys. \textbf{G8}, 1377 (1982).

\bibitem{IBM} F. Iachello and A. Arima, \textit{The Interacting Boson Model}
(Cambridge University Press, Cambridge, 1987).

\bibitem{GGG} H. Ganev, V. P. Garistov, and A. I. Georgieva,
Phys. Rev.  \textbf{C69}, 014305 (2004).

\bibitem{Van75} V. Vanagas, E. Nadjakov, and P. Raychev,
Preprint Trieste TC/75/40 (1975); V. Vanagas, E. Nadjakov, P.
Raychev, Bulg. J. Phys. \textbf{2}, 558-569 (1975).

\bibitem{TRM} N. Lo Iudice and F. Palumbo, Phys. Rev. Lett. \textbf{41}, 1532
(1978); Nucl. Phys. \textbf{A326}, 193 (1979).

\bibitem{Castanos87} O. Castanos, J. P. Draayer, and
Y. Leschber, Ann. Phys. \textbf{180}, 290 (1987).

\bibitem{Carvalho86} J. Carvalho \emph{et al.},
Nucl. Phys. \textbf{A452}, 240 (1986).

\bibitem{Nmin1} G. Rosensteel and D. J. Rowe, Phys. Rev. Lett. \textbf{46},
1119 (1981).

\bibitem{Nmin2} M. J. Carvalho, D. J. Rowe, S. Karram, and C. Bahri,
Nucl. Phys. \textbf{A703}, 167 (1987).

\bibitem{Nmin3} O. Castanos, P. O. Hess, J. P. Draayer, and P. Rochford, Nucl. Phys.
\textbf{A524}, 469 (1991).

\bibitem{Nmin4} D. Troltenier, J. P. Draayer, P. O. Hess, and O. Castanos, Nucl.
Phys. \textbf{A576}, 351 (1994).

\bibitem{Rowe85} D. J. Rowe, Rep. Prog. Phys. \textbf{48}, 1419
(1985).

\bibitem{Nilsson} S. G. Nilsson, Dan. Mat. Fys. Medd. \textbf{29}, 1 (1955).


\bibitem{pseudo-SU3} R. D. Ratnnu Raju, J. P. Draayer, and K. T.
Hecht, Nucl. Phys. \textbf{A202}, 433 (1973); J. P. Draayer and K.
J. Weeks, Ann. Phys. \textbf{156}, 41 (1984).

\bibitem{Van71} V. V. Vanagas,
\emph{Algebraic methods in nuclear theory} (Mintis, Vilnius, 1971)
(in russian).


\end{thebibliography}
\end{document}